\documentclass[aps,prd,twocolumn,floatfix,nofootinbib,superscriptaddress]{revtex4-1}
\usepackage{dcolumn}% Align table columns on decimal point
\usepackage{amsmath,amsfonts,extarrows,amssymb,mathrsfs,times,bm}
\usepackage{extarrows}
\usepackage{graphicx}
\usepackage{float}
\usepackage{graphicx,epstopdf}% Include figure files
\usepackage{dcolumn}% Align table columns on decimal point
\usepackage{exscale}
\usepackage{relsize}
\usepackage{mathtools}
\usepackage{mhchem}
\usepackage[colorlinks=true,breaklinks=true,linkcolor=blue,citecolor=blue,urlcolor=blue]{hyperref}
\newcommand{\nn}{\nonumber}

\begin{document}

\title{A Novel Method for Calculating Deflection Angle}
\author{Zonghai Li}
\email{sche-me@outlook.com}
\affiliation{Newton-Einstein academy, Jialidun University, Pu’er, 676213, China}

\date{\today}
\begin{abstract}

In this paper, we introduce a method for calculating the deflection angle in the weak-field approximation, applicable to both null and timelike rays. By combining the trajectory equation $\mathcal{Z}(u)=(du/d\phi)^2$ and the `straight line' $u(\varphi)={\sin\varphi}/b$, we introduce a new function $\Phi(\varphi)$. The deflection angle can then be expressed as $\delta=\Phi(0)+\Phi(\pi)-\pi$, which directly depends on the impact parameter rather than the closest approach distance. This method offers a convenient and straightforward approach to calculations, avoiding the complexities of integration or iterative procedures. As an illustrative application, we compute the deflection angle for charged particle in the Kerr-Newman spacetime.

\end{abstract}
\maketitle

\tableofcontents

\section{Introduction}

Due to the length of this paper, we will not discuss the history and significance of gravitational deflection and lensing, nor will we discuss the topics such as the successful detection of gravitational waves, the unveiling of black hole images, and so forth. However, we must add one more sentence: all of these are of great importance.

This paper focuses on the deflection angle in the weak-field approximation. Several methods have been developed for computing weak-field deflection angles. We primarily discuss two of them. The first is the direct integration method~\cite{Congdon-Keeton}, where the deflection angle is obtained by integrating the trajectory equation, as shown in Eq.~\eqref{IDA} below. This method has a long history and offers a straightforward physical interpretation, but the calculations can be difficult. In fact, many methods have been developed based on it to simplify the calculations, for example, Ref.~\cite{Jiaepjc}.

The second method, proposed by Gibbons and Werner~\cite{Gibbons-Werner,Werner2012}, involves applying the Gauss-Bonnet theorem to the optical metric space (or Jacobi metric space) to compute the deflection angle. This approach profoundly relates the deflection effect to the curvature and topology of space, and it has gained popularity among researchers in recent years (Please refer to the papers that cite Refs.~\cite{Gibbons-Werner,Werner2012} as their references). However, for computing higher-order deflection angle, this method requires iterative solutions, making it cumbersome and intricate~\cite{CGJ2019,LWJsd,LiJiaKD}.

In this paper, our goal is to introduce a novel method for computing weak-field deflection angles that significantly reduces computational complexity. Readers will see that this method can avoid challenging integrations and tedious iterations. This paper is structured as follows. In Sec.~\ref{I have traveled far and wide, roaming through the world's beautiful landscapes, yet it is you who still stirs my heart}, we present a novel method for calculating the deflection angle within the weak field approximation. Sec.~\ref{Each time I face the Chuan River, my emotions are unsettled} employs this method to compute the deflection angle of a charged particle in Kerr-Newman spacetime. Finally, our paper concludes briefly in Sec.~\ref{As time passes and seas dry and rocks crumble. suddenly, a thousand miles of sorrow dissolve in your presence}. The geometric unit is used throughout this paper.

\section{The method}
\label{I have traveled far and wide, roaming through the world's beautiful landscapes, yet it is you who still stirs my heart}

In Boyer-Lindquist coordinates $(t, r, \theta, \phi)$, the metric for the $4D$ stationary spacetime is given by
\begin{align}
\label{BL-metric}
	d s^2=&g_{t t}(r,\theta) d t^2+2 g_{t \phi}(r,\theta) d t d \phi+g_{rr}(r,\theta) dr^2\nn\\
	&+g_{\theta\theta}(r,\theta) d\theta^2+g_{\phi\phi}(r,\theta) d\phi^2.
\end{align}

This paper investigates the motion and deflection of particles on the equatorial plane $(\theta=\pi/2)$, with a specific emphasis on the weak-field approximation. When employing the variable $u=1/r$, the particle's trajectory equation can be expressed as follows
\begin{align}
\label{TED}
\mathcal{Z}(u)=\left(\frac{du}{d\phi}\right)^2.
\end{align}
The deflection angle can be determined from this trajectory equation as follows
\begin{align}
\label{IDA}
	\delta=2\int_0^{u_0} \frac{1}{\sqrt{\mathcal{Z}(u)}}du-\pi,
\end{align}
where $u_0=1/r_0$ and $r_0$ represents the closest approach distance. This equation may appear simple, but it is deceptively challenging.

We define a function $\Phi(u)$ as follows
\begin{align}
	\label{constant}
\pm\Phi(u)+C=\mp\int\frac{1}{\sqrt{\mathcal{Z}(u)}}du.
\end{align}
where $C$ is integration constant. By using the trajectory equation \eqref{TED}, we can derive
\begin{align}
	\label{Moliai1}
	\phi(u)=\begin{cases}
		-\Phi(u)+C & \text{if } |\phi(u)| <  |\phi(u_0)| \\
		\Phi(u)+C & \text{if } |\phi(u)| >  |\phi(u_0)|
	\end{cases},
\end{align}

Consider the following `straight line'
\begin{align}
	\label{FlatST}
	u=u(\varphi)=\frac{\sin\varphi}{b},\quad 0\leq\varphi\leq\pi.
\end{align}
Put it into Eq.~\eqref{Moliai1} to get the relationship between 
the angle of the particle ray and the angle of the ‘straight
line’, as follows
\begin{align}
	\label{Moliai2}
	\phi(\varphi)=\begin{cases}
		-\Phi(\varphi)+C & \text{if } \varphi < \frac{\pi}{2} \\
		\Phi(\varphi)+C & \text{if } \varphi >  \frac{\pi}{2}
	\end{cases},
\end{align}
where 
\begin{align}
	\label{Wangxiaobo}
	\pm\Phi(\varphi)+C=\mp\frac{1}{b}\int \frac{\cos\varphi}{\sqrt{\mathcal{Z}[u(\varphi)]}} d\varphi.
\end{align}

Note that $\Phi(\varphi)$ is not defined at $\varphi={\pi}/{2}$ since $u_0$ cannot be expressed using Eq.~\eqref{FlatST}. However, our main focus is on the boundaries, specifically $\varphi=0$ and $\varphi=\pi$. Assuming $\phi(0)<0$ and $\phi(\pi)>\pi$, we can express the deflection angle as
\begin{align}
	\delta=\left[\varphi-\phi(\varphi)\right]_{\varphi=0}+\left[\phi(\varphi)-\varphi\right]_{\varphi=\pi}.
\end{align}
Utilizing Eq.~\eqref{Moliai2}, the deflection angle becomes
\begin{align}
	\label{DAF}
	\delta=\Phi(0)+\Phi(\pi)-\pi.
\end{align}

The constant $C$ given by~\eqref{constant} indeed has no effect on the deflection angle. For convenience, we set $C=0$ and define $\Phi(\varphi)$ as follows
\begin{align}
	\label{Wang-er}
	\Phi(\varphi)=\left[\int \Theta(\varphi)~d\varphi\right]_{C=0},
\end{align}
where
\begin{align}
	\label{I wandered lonely as a cloud
		That floats on high o'er vales and hills,
		When all at once I saw a crowd,
		A host, of golden daffodils.}
	\Theta(\varphi)=-\frac{1}{b} \frac{\cos\varphi}{\sqrt{\mathcal{Z}[u(\varphi)]}}.
\end{align}

The deflection angle formula~\eqref{DAF} has several characteristics. First, it depends on the value of $\Phi(\varphi)$ at the boundary. Second, it uses the impact parameter instead of the closest approach distance. Eq.~\eqref{DAF} is very convenient for calculating the deflection angle in complex situations or high-order situations. In the next section, we will apply Eq.~\eqref{DAF} to address the complex scenario involving multiple parameters, specifically, the weak-field deflection angle of charged particle in a background of gravitational and electromagnetic fields, i.e., the Kerr-Newman spacetime. This model has previously been analyzed using the Gauss-Bonnet theorem in Ref.~\cite{LiJiaPRD2021}.

\section{Application to Kerr-Newman spacetime}
\label{Each time I face the Chuan River, my emotions are unsettled}

\subsection{Kerr-Newman metric}

The Kerr-Newman metric describes the spacetime outside a charged and rotating body with mass $M$, charge $Q$, and angular momentum $J=Ma$. In Boyer-Lindquist coordinates $(t, r, \theta, \phi)$, its line element is given by~\cite{KNJ}
\begin{align}
	\label{knmetric}
	ds^2=&-\left(1-\frac{2Mr-Q^2}{\Sigma}\right)dt^2+\frac{\Sigma}{\Delta}dr^2+\Sigma d\theta^2\nn\\
	&+\frac{1}{\Sigma}\left[\left(r^2+a^2\right)^2-\Delta a^2\sin^2\theta\right]\sin^2\theta d\phi^2\nn\\
	&-\frac{2a}{\Sigma}\left(2M r-Q^2\right)\sin^2\theta dt d\phi,
\end{align}

where
\begin{align}
	\Sigma=r^2+a^2\cos^2\theta,\quad\Delta=r^2-2Mr+a^2+Q^2.\nn
\end{align}

The corresponding gauge field reads
\begin{align}
	\label{gaugefield}
	A_\mu dx^\mu=-\frac{Qr}{\Sigma}\left(dt-a\sin^2\theta d\phi\right).
\end{align}

\subsection{The calculating of $\Phi(\varphi)$}

We consider the motion and deflection of a charged particle with mass $m$, energy $E$, and charge $q$ on the equatorial plane of the Kerr-Newman spacetime. The trajectory equation can, in principle, be derived from the geodesic equation. However, in this paper, we employ an alternative method, deriving it through the Jacobi metric. For detailed calculations, please refer to Appendix \ref{appJR} and Appendix \ref{appKNJR}. 

Consequently, the combined effects of the gravitational and electromagnetic fields on charged particle are described by the Kerr-Newman Jacobi-Randers metric given in Eq.~\eqref{KNJR}, and the particle's trajectory equation on the equatorial plane is given by Eq.~\eqref{chongsheng}. Following the path $Z[u(\phi)]\to \Theta(\varphi)\to \Phi(\varphi)$, we derive the function $\Phi(\varphi)$ using Eqs.~\eqref{I wandered lonely as a cloud
	That floats on high o'er vales and hills,
	When all at once I saw a crowd,
	A host, of golden daffodils.} and \eqref{Wang-er}, as presented in Eq.~\eqref{chuntianjiangxin}. In this expression, we approximate the result up to the second order in $\mathcal{O}(1/b)$. Since the deflection angle depends solely on the values of $\Phi(\varphi)$ at $\varphi=0$ and $\varphi=\pi$, we can omit terms that are obviously zero, such as those containing $\sin\varphi$. Additionally, we can set $\sqrt{\sec^2\varphi}=1$ since $\sqrt{\sec^2 0}=\sqrt{\sec^2 \pi}=1$. After implementing these two simplifications, we represent the remaining part as $\tilde{\Phi}(\varphi)$, given by
\begin{align}
\label{chuntian}
\tilde{\Phi}(\varphi)=\sum_{j=0}^2\tilde{\Phi}_j(\varphi)\left(\frac{1}{b}\right)^j+\mathcal{O}\left(\frac{1}{b^3}\right),
\end{align}
with
{\small
	\begin{align}
	\tilde{\Phi}_0=&-\varphi \cos \varphi,\quad \tilde{\Phi}_1=\frac{1}{v^2}\left[\left(1+v^2\cos^2\varphi\right)M-\frac{q}{E}Q\right],\nn\\
	\nn\\
	\tilde{\Phi}_2=&\left(\frac{q}{E}Q-2M\right)\frac{a}{v}+\frac{q}{E}\frac{M Q}{v^2}\left[3\varphi \cos\varphi\right] \nn\\
	&+\left[-3\left(\frac{1}{4}+\frac{1}{v^2}\right) \varphi\cos\varphi\right]M^2\nn\\
	&-\frac{1}{4}\bigg\{\frac{2q^2}{E^2 v^2}\left[\varphi \cos\varphi\right]-\left(1+\frac{2}{v^2}\right) \varphi\cos\varphi\bigg\}Q^2.\nn
\end{align}}
In the above, $v$ represents the asymptotic velocity of the particle (see Eq.~\eqref{LED}).

\subsection{Second-order deflection angle}

By utilizing Eq.~\eqref{chuntian} and taking advantage of the fact that ${\Phi}(0)=\tilde{\Phi}(0)$ and ${\Phi}(\pi)=\tilde{\Phi}(\pi)$, we can derive the following expressions
{\small\begin{align}
		\label{Jingdong0}
		\Phi(0)=&\frac{1}{b}\left[\left(1+\frac{1}{v^2}\right)M-\frac{q}{E v^2}Q\right]+\frac{a}{b^2v}\left(\frac{q}{E}Q-2M\right),\\
		\label{Jingdong1}
		\Phi(\pi)=&\frac{\pi}{4b^2}\left[3\left(1+\frac{4}{v^2}\right)M^2-\left(1+\frac{2}{v^2}-\frac{2q^2}{E^2v^2}\right)Q^2\right]\nn\\
		&+\pi+\Phi(0)-\frac{3\pi q}{Ev^2}QM.
\end{align}}
	
Subsequently, by substituting Eqs.~\eqref{Jingdong0} and \eqref{Jingdong1} into Eq.~\eqref{DAF}, we can derive the total second-order deflection angle as follows
{\small
\begin{align}
	\label{Result}
\delta=&2\frac{1}{b}\left[\left(1+\frac{1}{v^2}\right)M-\frac{q}{Ev^2}Q\right]+\frac{1}{b^2}\Bigg\{\nn\\
&\frac{\pi}{4}\left[3\left(1+\frac{4}{v^2}\right)M^2-\left(1+\frac{2}{v^2}-\frac{2q^2}{E^2v^2}\right)Q^2\right]\nn\\
&-3\pi\frac{q}{Ev^2}QM+\frac{2a}{v}\left(\frac{q}{E}Q-2M\right)\Bigg\}.
\end{align}
}

This result is exactly equivalent to the one obtained using the Gauss-Bonnet theorem, as presented in Eq.~(73) of Ref.\cite{LiJiaPRD2021}. The deflection angle \eqref{Result} exhibits some interesting characteristics. For instance, when $qQ/E=2M$, the parameter $a$ does not contribute to the deflection, up to a second-order approximation. From a geometric perspective, this is because in this scenario, the Kerr-Newman Jacobi-Randers metric \eqref{KNJR} has reversible geodesics. From a physical standpoint, this corresponds to a situation where the effects of the gravitomagnetic field and the magnetic field cancel each other out. For a more comprehensive discussion, please refer to Refs. \cite{LiJiaPRD2021} and \cite{LiJiaKD}.

\section{Conclusion}
\label{As time passes and seas dry and rocks crumble. suddenly, a thousand miles of sorrow dissolve in your presence}

This paper introduces a novel method for computing the weak-field deflection angle of particles moving on the equatorial plane in curved spacetime. By combining the particle's trajectory equation with the `straight line', we introduce a function $\Phi(\varphi)$ that characterizes the relationship between the angle of the particle ray and the angle of the `straight line'. The deflection angle can then be simply expressed as $\delta=\Phi(0)+\Phi(\pi)-\pi$. It entirely depends on the values of $\Phi(\varphi)$ at the boundaries and is directly characterized by the impact parameter $b$ rather than the closest approach distance $r_0$. 

The primary advantage of this method lies in its computational simplicity: it does not involve challenging integrals, particle trajectory solving, or tedious iterations. As a straightforward application, we compute the deflection angle of charged particle in the Kerr-Newman spacetime, and the result is in complete agreement with existing literature. Due to space constraints, we cannot demonstrate the power of using our method to compute higher-order deflection angles in this paper.

We anticipate that this method can be extended to various scenarios, such as computing finite-distance deflection angles, where it is assumed that the distances from the particle source to the lens and from the observer to the lens are finite. This scenario has already been effectively addressed using the Gauss-Bonnet theorem~\cite{ISOA2016,OIA2017}. Certainly, the extension to the finite-distance case is straightforward ~\cite{Li-Finite}. We also hope that the ideas included in this method can provide insights into other directions.

%\acknowledgements

\appendix

\section{Jacobi-Randers metric and trajectory equation}
\label{appJR}

Let's examine the gravitational field $g_{\mu\nu}$ and electromagnetic field $A_{\mu}$ in the coordinates $(t,x^i)~(i=1,2,3)$
\begin{align}
&d s^2=g_{t t}(x) d t^2+2 g_{t i}(x) d t d x^i+g_{i j}(x) d x^i d x^j,\\
&A_{\mu}dx^{\mu}=A_{t}(x)dt+A_i(x)dx^i.
\end{align}

Now, let's consider the motion of a charged particle with mass $m$, energy $E$, and charge $q$ under the influence of both the gravitational field and the electromagnetic field. Due to the conservation of the particle's energy in this scenario, its trajectory corresponds to the geodesic of the $3D$ Jacobi metric. The Jacobi metric is given by~\cite{Chanda-Jacobi}
\begin{align}
	\label{Jacobi-Randers}
F(x,dx)=\sqrt{\alpha_{ij}dx^idx^j}+\beta_idx^i,
\end{align}
where
\begin{align}
\label{fensleranders1}
&\alpha_{ij}=\frac{\left(E+qA_t\right)^2+m^2{g}_{tt}}{-{g}_{tt}}\left({g}_{ij}-\frac{{g}_{ti}{g}_{tj}}{{g}_{tt}}\right),\\
\label{fensleranders2}
&\beta_i=qA_i-\left(E+qA_t\right)\frac{g_{ti}}{g_{tt}}.
\end{align}

The Jacobi metric \eqref{Jacobi-Randers} is a Randers type Finsler metric, where $\alpha_{ij}$ is the Riemannian metric, and $\beta_i$ is a one-form, satisfying $\sqrt{\alpha^{ij}\beta_i\beta_j}<1$. For convenience, we refer to it as the Jacobi-Randers metric.

Now, let's consider Boyer-Lindquist coordinates $(x^1=r, x^2=\theta, x^3=\phi)$ and suppose $g_{\mu\nu}=g_{\mu\nu}(r, \theta)$ and $A_{\mu}=A_{\mu}(r, \theta)$. In this case, the particle possesses a conserved angular momentum $L$. On the equatorial plane, where $\theta=\pi/2$ and $x=(r, \phi)$, the orbital equation of the particle can be expressed in terms of $(\alpha_{ij}, \beta_i)$ as follows~\cite{LWJsd}
\begin{align}
	\label{orbitequation}
	\mathcal{Z}(u)=\left(\frac{du}{d\phi}\right)^2=u^4\frac{\alpha_{\phi\phi}\left[\alpha_{\phi\phi}-\left(L-\beta_\phi\right)^2\right]}{\alpha_{rr}\left(L-\beta_\phi\right)^2},
\end{align}
where $u=1/r$.

The energy $E$ and angular momentum $L$ are commonly expressed in terms of the particle's asymptotic velocity $v$ and the impact parameter $b$ as follows 
\begin{align}
	\label{LED}
	E=\frac{m}{\sqrt{1-v^2}},\quad L=bvE=\frac{mbv}{\sqrt{1-v^2}}.
\end{align}

\section{Kerr-Newman Jacobi-Randers metric and $\Phi(\varphi)$}
\label{appKNJR}

Substituting the Kerr-Newman metric~\eqref{knmetric} and its electromagnetic field~\eqref{gaugefield} into~\eqref{Jacobi-Randers}, and considering the equatorial plane scenario, we obtain the following $2D$ Jacobi-Randers metric
\begin{align}
		\label{KNJR}
		F(r,\phi,dr,d\phi)=\sqrt{\alpha_{rr}dr^2+\alpha_{\phi\phi} d\phi^{2}}+\beta_{\phi}d\phi,
\end{align}
where
\begin{align}
\label{com-KNJR1}
\alpha_{rr}=&\left[\frac{\left(q Q-Er\right)^{2}}{\Delta-a^2}-m^{2}\right]\frac{r^2}{\Delta},\\ 
\label{com-KNJR2}
\alpha_{\phi\phi}=&\left[\frac{\left(q Q-Er\right)^{2}}{\Delta-a^2}-m^{2}\right]\frac{r^2\Delta}{\Delta-a^2},\\
\label{com-KNJR3}
\beta_{\phi}=&\frac{a\left[E\left(Q^{2}-2 M r\right)+q Q r\right]}{\Delta-a^2}.
\end{align}

Putting the Kerr-Newman Jacobi-Randers data \eqref{com-KNJR1}-\eqref{com-KNJR3} into the trajectory equation \eqref{orbitequation}, and using Eqs.~\eqref{FlatST} and \eqref{LED}, we get
{\small
\begin{align}
	\label{chongsheng}
\mathcal{Z}(\varphi)=&\frac{1}{b^2} \cos^2\varphi+2\left[\left(1-v^2 \cos^2 \varphi\right) M-\frac{q Q}{E}\right] \frac{\sin\varphi}{b^3 v^2}\nn\\
	&+\left[\frac{2}{v}\left(\frac{q Q}{E}-2 M\right) +{\left(2+\cos2\varphi\right)a\sin\varphi}\right]\frac{a \sin\varphi}{b^4}\nn\\
	&-\left(1-v^2 \cos^2\varphi-\frac{q^2}{E^2}\right) \frac{Q^2 \sin^2\varphi}{b^4 v^2}+\mathcal{O}\left(\frac{\epsilon^3}{b^5}\right),
\end{align}
}
where $\epsilon\in\{M,a,Q\}$.

Utilizing Eq.~\eqref{chongsheng}, we can compute $\Theta(\varphi)$ as defined in Eq.\eqref{I wandered lonely as a cloud That floats on high o'er vales and hills, When all at once I saw a crowd, A host, of golden daffodils.}, and the outcome is approximated up to the second order in $\mathcal{O}(1/b)$ as
\begin{align}
	\label{jiangxinchuntian}
		\Theta(\varphi)=\sum_{j=0}^2\Theta_j(\varphi)\sqrt{\sec^2\varphi}\left(\frac{1}{b}\right)^j+\mathcal{O}\left(\frac{1}{b^3}\right),
\end{align}
where
{\small
	\begin{align}
		\Theta_0=&-\cos \varphi,~~\Theta_1=\left[\left(1-v^2 \cos^2\varphi\right) M-\frac{q}{E} Q\right]\frac{\tan\varphi}{v^2},\nn\\
		\nn\\
		\Theta_2=
			& \bigg[\left(\frac{q Q}{E}-2 M\right) \frac{a}{v}+\frac{a^2}{2}\left(1+2 \cos^2\varphi\right) \sin\varphi\nn\\
			&-\frac{Q^2}{2 v^2}\left(1-v^2 \cos^2\varphi-\frac{q^2}{E^2}+\frac{3 q^2 \sec^2\varphi }{E^2v^2}\right)\sin\varphi\nn\\
			&~~~~~~~-\frac{3M^2}{2v^4}\left(1-v^2 \cos^2\varphi+\frac{2q Q }{EM }\right)\nn\\
			&~~~~~~~~~~~~~~~~~\times\left(1-v^2 \cos^2\varphi\right)\sec\varphi\tan\varphi\bigg]{\tan\varphi}.\nn
	\end{align}
}

By substituting the newly obtained $\Theta(\varphi)$ into Eq.~\eqref{Wang-er}, we can derive $\Phi(\varphi)$ as follows
	\begin{align}
		\label{chuntianjiangxin}
		\Phi(\varphi)=\sum_{j=0}^2\Phi_j(\varphi)\sqrt{\sec^2\varphi}\left(\frac{1}{b}\right)^j+\mathcal{O}(\frac{1}{b^3}),
    \end{align}
where
{\small
	\begin{align}
		\Phi_0=&-\varphi \cos \varphi,\quad \Phi_1=\frac{1}{v^2}\left[\left(1+v^2\cos^2\varphi\right)M-\frac{q}{E}Q\right],\nn\\
		\nn\\
		\Phi_2=&\left(\frac{q}{E}Q-2M\right)\frac{a}{v}+\frac{1}{2}a^2\sin^3\varphi+\frac{q}{E}\frac{M Q}{v^2}\left[3\varphi \cos\varphi\right.\nn\\
		&\left.-\sin\varphi\left(3-\frac{\tan^2\varphi}{v^2}\right)\right] -\frac{1}{4}\bigg\{\frac{2q^2}{E^2 v^2}\left[\varphi \cos\varphi\right.\nn\\
		&\left.-\sin\varphi\left(1-\frac{\tan^2\varphi}{v^2}\right)\right]-\left(1+\frac{2}{v^2}\right) \varphi\cos\varphi\nn\\
		&+\left(\frac{2}{v^2}+\cos^2\varphi\right) \sin\varphi\bigg\}Q^2+\left[-3\left(\frac{1}{4}+\frac{1}{v^2}\right) \varphi\cos\varphi\right.\nn\\
		&\left.+\left(\frac{1}{2 v^4}+\frac{3}{v^2}+\frac{3 \cos^2\varphi}{4}-\frac{\sec^2\varphi}{2 v^4}\right) \sin\varphi\right]M^2.\nn
	\end{align}
}

Note that, while the $M^2$ term is not present in $\mathcal{Z}(\varphi)$ (see Eq.~\eqref{chongsheng}), it does appear in both $\Theta(\varphi)$ (see Eq.~\eqref{jiangxinchuntian}) and $\Phi(\varphi)$ (see Eq.~\eqref{chuntianjiangxin}). Consequently, the second-order deflection angle, as presented in Eq.~\eqref{Result}, includes the $M^2$ term.

\end{document}